 \documentclass[12pt]{article}
\usepackage{fullpage}
\usepackage{amssymb}
\usepackage{amsmath}
\usepackage{bm}

\usepackage{mathptmx}       % selects Times Roman as basic font
\usepackage{helvet}         % selects Helvetica as sans-serif font
\usepackage{courier}        % selects Courier as typewriter font
\usepackage{makeidx}         % allows index generation
\usepackage{graphicx}        % standard LaTeX graphics tool
\usepackage{multicol}        % used for the two-column index
\usepackage{wrapfig}
\usepackage{color}
\usepackage{cite}

\def\magenta{\textcolor{magenta}}

 \def\eps{\varepsilon}

 \def\epso{\eps_{\scriptscriptstyle 0}}

\def\muo{\mu_{\scriptscriptstyle 0}}
\def\ko{k_{\scriptscriptstyle 0}}

\def\##1{{\boldsymbol{#1}}}
\def\=#1{\underline{\underline #1}}
\def\~#1{\tilde{#1}}
\def\^#1{\breve{#1}}
\def\`#1{{#1^\prime}}
\def\:#1{#1^{\prime\prime}}

\def\les{\left[}
\def\ris{\right]}
\def\lec{\left\{}
\def\ric{\right\}}

\def\l#1{\label{#1}}
\def\r#1{(\ref{#1})}

\def\un{\hat{\#n}}

\def\.{\mbox{ \tiny{$^\bullet$} }}

\def\sca{_{\rm sca}}
\def\inside{_{\rm int}}
\def\source{_{\rm source}}
\def\body{_{\rm obj}}
\def\xs{_{\rm xs}}

\DeclareMathAlphabet{\mathpzc}{OT1}{pzc}{m}{it}
 \def\calV{{\mathpzc V}}
\def\calS{{\mathpzc S}}

 \def\calVi{{\mathpzc V}_{\rm i}}
  \def\calVe{{\mathpzc V}_{\rm e}}
  \def\calVs{{\mathpzc V}_{\rm s}}

\def\nuee{\={\nu}_{ee}(\#r)}
\def\nuem{\=\nu_{em}(\#r)}
\def\nume{\=\nu_{me}(\#r)}
\def\numm{\=\nu_{mm}(\#r)}

\def\nueep{\=\nu_{ee}(\#r^\prime)}
\def\nuemp{\=\nu_{em}(\#r^\prime)}
\def\numep{\=\nu_{me}(\#r^\prime)}
\def\nummp{\=\nu_{mm}(\#r^\prime)}

\def\plus{_{+}}
\def\minus{_{-}}

\def\Lplus{\lec\nabla\times\=I+i\omega\les\=\xi+(\#K-\#\Gamma)\times\=I\ris\ric}
\def\Lminus{\lec\nabla\times\=I+i\omega\les\=\xi-(\#K+\#\Gamma)\times\=I\ris\ric}

\def\lplus{\={{\mathfrak{L}}}\plus}
\def\lminus{\={{\mathfrak{L}}}\minus}

\begin{document}

\magenta{Accepted for publication in: Journal of Optics (Springer Nature)}\\

\begin{center}
\textbf{New Principle for Scattering Inside A Huygens   Bianisotropic Medium}\\[10pt]
\textit{Akhlesh Lakhtakia}\\[5pt]
ORCID: 0000-0002-2179-2313\\[5pt]
{Nanoengineered Metamediums Group,  Department of
Engineering Science and Mechanics,  Pennsylvania State University,
University Park, PA  16802, USA
\\  \textit{akhlesh@psu.edu}}

\date{\today}
\end{center}
 
\noindent \textbf{Abstract} A Huygens bianisotropic medium 
is a linear homogeneous  medium for which the Huygens principle can be formulated.
When a  bounded 3D scattering object composed of a linear bianisotropic medium, whether homogeneous or
not,
is embedded in a Huygens bianisotropic medium,
the excess field phasors inside that object 
act  as volume current densities, and the tangential components of the internal field phasors on the
surface of the same object act  as surface current densities, to radiate identical field phasors in the external region.
 
\vspace{5mm}

\section{Introduction}
Frequency-domain scattering by a bounded three-dimensional (3D) object in free space can be handled
theoretically by analytical, semi-analytical, and numerical methods, depending on the shape and
the composition of the
object. Perhaps the simplest case is of a sphere composed of a homogeneous isotropic medium \cite{Kerker1},
whether the medium is dielectric-magnetic \cite{Lorenz,Mie,BH83} or biisotropic \cite{Bohren1974,Beltrami}.
Analytical solutions  based on a separation-of-variables approach are available
Scattering by a spheroid composed of a homogeneous medium of one of those types has also been formulated
and solved analytically \cite{Asano,Cooray}. The same approach can be extended to radially stratified spheres \cite{Bhandari,Jaggard1999} and
confocally stratified spheroids  made of  isotropic and biisotropic mediums. 

Except for special cases \cite{Monzon,Qiu,Jafri}, scattering
by
spheres and spheroids made of homogeneous anisotropic mediums requires the use of semi-analytical \cite{Kiselev,Li2012,Novitsky2017,Alkhoori3,Alkhoori2}
and numerical
methods \cite{Varadan1989,Sadati,Alkhoori3}. 
 
The extended
boundary condition method is semi-analytical. It requires  bilinear expansions of the infinite-medium
dyadic Green functions for the medium of which the 3D object is composed  \cite{L-Magdy}. This requirement has restricted the conventional formulation of this method to objects composed of  a homogeneous medium that  either are
biisotropic \cite{LVV-AO}   or belong to a certain class of bianisotropic mediums \cite{LM-joi,Alkhoori3}. 
However, that requirement can be bypassed   for a 3D object made of a homogeneous orthorhombic medium with \cite{Zouros2021,Geng2012} or without \cite{Schmidt2009} gyrotropy, because basis functions to represent the 
actual electric and magnetic field phasors   inside the object can be synthesized from an angular spectrum of plane waves. Of course, the formulation and solution processes then become considerably toilsome.

Numerical methods can handle arbitrarily shaped 3D objects that are composed of a homogeneous/nonhomogeneous
and isotropic/biisotropic/anisotropic/bianisotropic medium. The finite-element method \cite{Ishii2014,Yang2021}
and the method of moments \cite{Strong,Mei,Maddah-Ali} partition the scattering object
into several subregions.  Suitable  basis functions are then used
to represent the electric and magnetic field phasors  inside each subregion. The discrete dipole approximation partitions the
scattering object into multiple subobjects.  Each subobject
is modeled  by   a set of  three mutually orthogonal  electric and three mutually orthogonal magnetic dipoles of strengths and phases to be determined   \cite{Strong,OSA}. 
Appropriate only for  homogeneous objects, the boundary element method partitions the object's surface and uses subregional basis functions
 to determine the actual electric and magnetic field phasors on the object's surface \cite{Yla,Sun,Cui}.
 These and other numerical methods \cite{Kahnert}
 are also hybridized for faster and/or higher-resolution computations  \cite{Zhu2013,Liu2019,Yang2021}. Finally, the finite-difference time-domain method solves the differential form of the Maxwell equations directly on a spacetime grid that encompasses the scattering object
\cite{Prokopidis,Gansen}.

The  absence of purely analytical methods to treat scattering by  3D objects of arbitrary shape and composition
highlights the need for computationally tractable principles that the results obtained from every semi-analytical and numerical method must satisfy. One such principle is  mathematical, concerned
with  the completeness and convergence of  representations of the scattered and the internal field phasors \cite{Kahnert,Massoudi,Aydin}. 
The second principle is physical, that of conservation of energy   in any time-invariant system \cite{Noether}. This principle
is often enforced through
the optical theorem \cite{deHoop,Newton}. 
Energy, however, is a   quantity derived from   electric and magnetic fields. For frequency-domain scattering problems, a universal principle involving  the electric and magnetic field phasors directly is desirable.

In this paper, two new relations are obtained for 3D frequency-domain scattering problems involving:
\begin{itemize}
\item[(i)] the actual electric and magnetic field phasors   inside the  object,
\item[(ii)] the actual electric and magnetic field phasors on the surface of the  object, and
\item[(iii)] the infinite-medium dyadic Green functions \cite{DGFbook} of the linear
homogeneous medium in which the object is embedded.
\end{itemize}
The external medium is assumed to extend to infinity in all directions. 
This medium  need not be free space. Instead, it can be the
most general linear, homogeneous, bianisotropic medium for 
which the Huygens principle has been formulated \cite{FL-PLA}. We use the phrasal adjective \textit{Huygens bianisotropic}
for such a medium. 
The internal medium is also linear; it can be
homogeneous/non\-homo\-geneous
and isotropic/biisotropic/aniso\-tropic/bianisotropic.  
The two new  relations, which can be derived from one another by using the frequency-domain Maxwell equations
applied to the external medium, constitute a principle that every applicable semi-analytical and numerical method must satisfy. This new principle is independent of the principle of conservation of energy.  

The plan of the paper is as follows. Section~\ref{theory} sets up the 3D scattering problem by providing the constitutive relations of the external and internal mediums, the governing differential equations, and the infinite-medium dyadic Green functions  of the external medium. Two volume integral equations are formulated in Sec.~\ref{vie} and two
surface integral equations are presented in Sec.~\ref{sie}. Comparisons of the volume and surface integral equations deliver two new relations in Sec.~\ref{newrels}. Finally, in Sec.~\ref{newprin},
these relations underwrite the new principle for scattering inside   any Huygens bianisotropic
 medium.

An $\exp(-i\omega t)$ time-dependence is assumed with $\omega$ as the angular frequency, $t$  the time, and  $i=\sqrt{-1}$.   Single-underlined letters denoted vectors, with $\#0$ denoting the null vector and the caret ($\hat{}$) identifying unit vectors.  Dyadics are underlined twice, with $\=I$ denoting the identity dyadic and $\=0$ the null dyadic.

\section{Theoretical Preliminaries}\label{theory}

\begin{figure}[ht!]
\begin{center}
\includegraphics[width=0.4\textwidth]{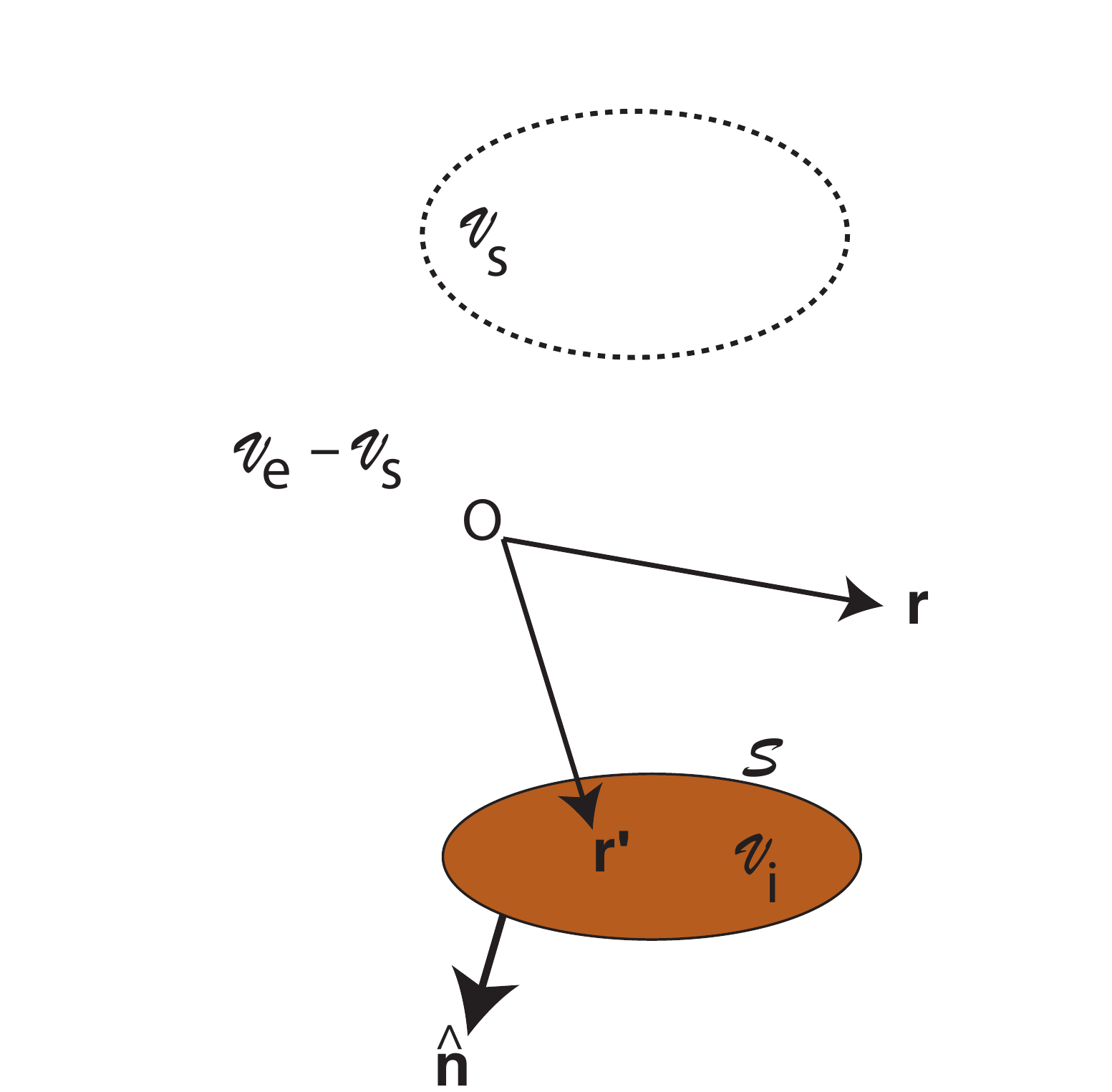} 
\caption{Schematic showing all space $\calV$ divided  into the  external region ${\calVe}$
and the  internal region ${\calVi}$
separated by the closed surface $\calS$. The  source of the electromagnetic field is confined to the 
region ${\calVs}\subset{\calVe}$. 
 \label{Fig1}}
\end{center}
\end{figure}

Suppose that all space $\calV$ is divided into 
the unbounded region ${\calVe}$ and the bounded region ${\calVi}$, the boundary of the two regions
being the closed
surface  $\calS$,
as shown in Fig.~\ref{Fig1}. The external region ${\calVe}$ is
occupied by a Huygens bianisotropic medium   characterized in the frequency domain  by 
the linear constitutive relations \cite{L-Magdy,FL-PLA}
 \begin{equation}
\left.\begin{array}{l}
\#D(\#r)=\=\eps \.\#E(\#r)+\les
{\=\xi}+ \left(\#K-\#\Gamma\right) \times\=I\ris
\.\#H(\#r)\\
[5pt]
\#B(\#r)=\=\mu \.\#H(\#r)-\les
{\=\xi}
- \left(\#K+\#\Gamma\right)\times\=I\ris
 \.\#E(\#r)
 \end{array}\right\}
 \,,\qquad \#r\in{\calVe}\,.
 \label{con2or-e}
\end{equation}
Here and hereafter,  $\#r$ is the position vector, $\#E(\#r)$   the electric field phasor, $\#H(\#r)$   the magnetic field
phasor, $\#D(\#r)$   the electric displacement phasor, and $\#B(\#r)$   the magnetic induction phasor.
The arbitrary vectors  $\#K$ and $\#\Gamma$ as well as
the  symmetric dyadics
$\=\eps=\=\eps ^{\rm T}$,
$\=\mu= \=\mu ^{\rm T}$, and
${\=\xi}
={\=\xi}
^{\rm T}$
are implicit functions of   $\omega$,
 and the superscript $^{\rm T}$ denotes the transpose. 
  
 %%%%%%%%%% Fig 1 begins %%%%%%%%%%%%% 
%%%%%%%%%% Fig 1 ends %%%%%%%%%%%%%

A linear bianisotropic medium also occupies the internal region ${\calVi}$, but it need not be homogeneous.
Its frequency-domain constitutive relations  are stated as
 \begin{equation}
\left.\begin{array}{l}
\#D(\#r)=
\les\=\eps+\nuee\ris \.\#E(\#r)+\les
{\=\xi}+ \left(\#K-\#\Gamma\right) \times\=I+\nuem\ris
\.\#H(\#r)\\
[5pt]
\#B(\#r)=
\les\=\mu+\numm\ris \.\#H(\#r)-\les
{\=\xi}
- \left(\#K+\#\Gamma\right)\times\=I-\nume\ris
 \.\#E(\#r)
 \end{array}\right\}
 \,,\qquad \#r\in{\calVi}\,.
 \label{con2or-i}
\end{equation}
No constraints are placed on the dyadic functions
$\nuee$, etc.
Thus, the internal medium can be the most general linear bianisotropic medium.

The  primary source of the electromagnetic field phasors is confined to the bounded
region $\calV_s\subset{\calVe}$.   Therefore,
the source electric current density phasor  $\#J_e (\#r)$  and  
  the source magnetic current density phasor  $\#J_m (\#r)$ 
  are such that
  \begin{equation}
\left.\begin{array}{l}
\#J_e (\#r) =\#0
\\[5pt]
\#J_m (\#r) =\#0
\end{array}
\right\}\,,\qquad \#r\notin\calV_s\,.
\end{equation}
It is assumed that $\calV_s$ is situated so far from $\calVi$ that $\#J_e (\#r)$ and $\#J_m (\#r)$
are negligibly affected by the scattered field phasors.

 \subsection{Governing differential equations}

The electromagnetic field phasors  everywhere  satisfy the frequency-domain Maxwell curl postulates
\begin{equation}
\left.\begin{array}{l}
\nabla\times\#H(\#r)+i\omega\#D(\#r)=\#J_e(\#r) 
\\[5pt]
\nabla\times\#E(\#r)-i\omega\#B(\#r)=-\#J_m(\#r) 
\end{array}\right\}\,,\qquad\#r\in \calV\,.
\label{M1worM2wor-0}
\end{equation}
Substitution
of  Eqs.~\r{con2or-e} and \r{con2or-i}  in Eqs.~\r{M1worM2wor-0} yields
\begin{equation}
\left.\begin{array}{l}
\lplus\.\#H(\#r)+i\omega
\=\eps \.\#E(\#r)=\#J_e(\#r) 
\\[5pt]
\qquad\qquad
-i\omega\nuee\.\#E(\#r) -i\omega\nuem\.\#H(\#r)
\\[5pt]
\lminus\.\#E(\#r)-i\omega
\=\mu \.\#H(\#r)
=-\#J_m(\#r)  
\\[5pt]
\qquad\qquad
+ i\omega\nume\.\#E(\#r) +i\omega\numm\.\#H(\#r)
\end{array}\right\}\,,\qquad\#r\in \calV\,,
\label{M1worM2wor}
\end{equation}
where the operators
 \begin{equation}
 \left.\begin{array}{l}
 \lplus=\Lplus
 \\[5pt]
 \lminus=\Lminus
 \end{array}
 \right\}\,
 \end{equation}
 and
 the dyadics
 \begin{equation}
\left.\begin{array}{l}
\nuee =\#0
\\[5pt]
\nume =\#0
\\[5pt]
\nuem =\#0
\\[5pt]
\numm =\#0
\end{array}
\right\}\,,\qquad \#r\in{\calVe}\,.
\end{equation}
 Equations~\r{M1worM2wor} govern the electric and magnetic field phasors throughout $\cal V$.

\subsection{Dyadic Green functions}
  
Four  infinite-medium dyadic Green functions $\=G^{pq}(\#r,\#r^\prime)$, $p\in\lec{e,m}\ric$
and $q\in\lec{e,m}\ric$,
 can be prescribed for the chosen external
medium \cite[Sec.~1.3.1]{DGFbook}. Of these,
$\=G^{ee}(\#r,\#r^\prime)$ and $\=G^{mm}(\#r,\#r^\prime)$ are, respectively, the solutions of
the differential equations
\begin{equation}
\left(\lplus\.\=\mu^{-1}\.\lminus
-\omega^2\=\eps\right)
\.\=G^{ee}(\#r,\#r^\prime)=i\omega\=I\delta(\#r-\#r^\prime)
\label{six}
\end{equation}
and
\begin{equation}
\left(\lminus\.\=\eps^{-1}\.\lplus
-\omega^2\=\mu\right)
\.\=G^{mm}(\#r,\#r^\prime)=i\omega\=I\delta(\#r-\#r^\prime)
\,,
\label{seven}
\end{equation}
where $\delta(\.)$ is the Dirac delta, $\#r$ serves as the field
point, and $\#r^\prime$ is the source point. The remaining two
dyadic Green functions  can be obtained from the first two
as  
\begin{equation}
\displaystyle{
\=G^{em}(\#r,\#r^\prime)=-\frac{1}{i\omega}\,\=\eps^{-1} \.
\lplus
\.\=G^{mm}(\#r,\#r^\prime)
}
\label{five-em}
\end{equation}
and
\begin{equation}
\displaystyle{
\=G^{me}(\#r,\#r^\prime)=\frac{1}{i\omega}\,\=\mu^{-1} \.
\lminus
\.\=G^{ee}(\#r,\#r^\prime) 
}\,.
\label{five-me}
\end{equation}
These four functions have the following symmetries
with respect to the interchange of the source and field points \cite{L-Magdy}:
 \begin{equation}
\left.\begin{array}{l}
\=G^{ee}(\#r,\#r^\prime)= \lec\=G^{ee}(\#r^\prime,\#r)\ric^{\rm T}
\exp\les2i\omega\#\Gamma\.\left(\#r-\#r^\prime\right)\ris
\\[5pt]
\=G^{mm}(\#r,\#r^\prime)= \lec\=G^{mm}(\#r^\prime,\#r)\ric^{\rm T}
\exp\les2i\omega\#\Gamma\.\left(\#r-\#r^\prime\right)\ris
\\[5pt]
\=G^{em}(\#r,\#r^\prime)= -\lec\=G^{me}(\#r^\prime,\#r)\ric^{\rm T}
\exp\les2i\omega\#\Gamma\.\left(\#r-\#r^\prime\right)\ris
\end{array}\right\}\,.
\label{DGF-symm}
\end{equation}

Closed-form expressions for $\=G^{pq}(\#r,\#r^\prime)$, $p\in\lec{e,m}\ric$
and $q\in\lec{e,m}\ric$, are not available for every   medium described
by Eqs.~\r{con2or-e}. However, $\=G^{pq}(\#r,\#r^\prime)$ can always be synthesized
using the spatial Fourier transform \cite[Sec.~1.4.1]{DGFbook}.

\section{Volume integral equations}\label{vie}
The solutions of Eqs.~\r{M1worM2wor}  come in two parts.
The first part is due to the source current density phasors in $\calV_s$, the second
due to the scattering object constituted. 
Accordingly, the \textit{actual} field phasors everywhere are written as
\begin{equation}
\left.\begin{array}{l}
\#E(\#r)=\#E\source(\#r)+\#E\body(\#r)
\\[5pt]
\#H(\#r)=\#H\source(\#r)+\#H\body(\#r)
\end{array}
\right\}\,,\quad \#r\in\calV\,.
\label{GeorGmor}
\end{equation}

By definition, the field phasors
\begin{equation}
\displaystyle{
\#E\source(\#r)=\int\limits_{{\calV_s}}
\les \=G^{ee}(\#r,\#r^\prime)\.\#J_e(\#r^\prime)+\=G^{em}(\#r,\#r^\prime)\.\#J_m(\#r^\prime)\ris d^3\#r^\prime
}\,,\quad \#r\in \calV\,,
\label{GeorGmor-es}
\end{equation}
and
\begin{equation}
\displaystyle{
\#H\source(\#r)=\int\limits_{{\calV_s}}
\les \=G^{me}(\#r,\#r^\prime)\.\#J_e(\#r^\prime)+\=G^{mm}(\#r,\#r^\prime)\.\#J_m(\#r^\prime)\ris d^3\#r^\prime
}\,, \,\,\#r\in \calV\,,
\label{GeorGmor-ms}
\end{equation}
exist everywhere when   ${\calVi}$ is \textit{also} occupied by the external medium.
Analogously,
the field phasors
\begin{eqnarray}
 \nonumber
&&
\displaystyle{
\#E\body(\#r)= -i\omega\int\limits_{{{\calVi}}} 
\lec \=G^{ee}(\#r,\#r^\prime)\.\les \nueep\.\#E(\#r^\prime) +\nuemp\.\#H(\#r^\prime)\ris
\ric d^3\#r^\prime
}
\\[5pt]
&&
\displaystyle{
-i\omega\int\limits_{{{\calVi}}}
\lec
\=G^{em}(\#r,\#r^\prime)\. \les\numep\.\#E(\#r^\prime) + \nummp\.\#H(\#r^\prime)\ris
\ric d^3\#r^\prime
}\,,\,\, \#r\in\calV\,,
\label{GeorGmor-eb}
\end{eqnarray}
and
\begin{eqnarray}
 \nonumber
&&
\displaystyle{
\#H\body(\#r)= -i\omega\int\limits_{{{\calVi}}}
\lec \=G^{me}(\#r,\#r^\prime)\.\les \nueep\.\#E(\#r^\prime) +\nuemp\.\#H(\#r^\prime)\ris
\ric d^3\#r^\prime
}
\\[5pt]
&&
\displaystyle{
-i\omega\int\limits_{{{\calVi}}}
\lec
\=G^{mm}(\#r,\#r^\prime)\. \les\numep\.\#E(\#r^\prime) + \nummp\.\#H(\#r^\prime)\ris
\ric d^3\#r^\prime
}\,,\,\, \#r\in\calV\,,
\label{GeorGmor-mb}
\end{eqnarray}
come into existence  only if the internal medium   is different from the external medium.
Thus, the \textit{excess} field phasors
 \begin{equation}
\left.\begin{array}{l}
\#D\xs(\#r)=
 \nuee  \.\#E(\#r)+ \nuem 
\.\#H(\#r)\\
[5pt]
\#B\xs(\#r)=
\numm  \.\#H(\#r)+\nume 
 \.\#E(\#r)
 \end{array}\right\}
 \,,\qquad \#r\in{\calVi}\,,
 \label{excess}
\end{equation}
are   \textit{equivalent} to
volume current densities  in ${\calVi}$.

When Eqs.~\r{GeorGmor-eb} and \r{GeorGmor-mb} are substituted in Eqs.~\r{GeorGmor}, the resulting
equations
\begin{eqnarray}
 \nonumber
&&
\displaystyle{
\#E(\#r)= \#E\source(\#r)-i\omega\int\limits_{{{\calVi}}} 
\lec \=G^{ee}(\#r,\#r^\prime)\.\les \nueep\.\#E(\#r^\prime) +\nuemp\.\#H(\#r^\prime)\ris
\ric d^3\#r^\prime
}
\\[5pt]
&&
\displaystyle{
-i\omega\int\limits_{{{\calVi}}}
\lec
\=G^{em}(\#r,\#r^\prime)\. \les\numep\.\#E(\#r^\prime) + \nummp\.\#H(\#r^\prime)\ris
\ric d^3\#r^\prime
}\,,\,\, \#r\in\calV\,,
\label{GeorGmor-e}
\end{eqnarray}
and
\begin{eqnarray}
 \nonumber
&&
\displaystyle{
\#H (\#r)=\#H\source(\#r) -i\omega\int\limits_{{{\calVi}}}
\lec \=G^{me}(\#r,\#r^\prime)\.\les \nueep\.\#E(\#r^\prime) +\nuemp\.\#H(\#r^\prime)\ris
\ric d^3\#r^\prime
}
\\[5pt]
&&
\displaystyle{
-i\omega\int\limits_{{{\calVi}}}
\lec
\=G^{mm}(\#r,\#r^\prime)\. \les\numep\.\#E(\#r^\prime) + \nummp\.\#H(\#r^\prime)\ris
\ric d^3\#r^\prime
}\,,\,\, \#r\in\calV\,,
\label{GeorGmor-m}
\end{eqnarray}
are {\it volume integral equations}  because the unknown quantities,
 $\#E (\#r)$ and $\#H (\#r)$, appear inside as well as outside the volume integrals.

\section{Surface integral equations}\label{sie}
Mathematical statements of the Huygens principle pertinent to the region ${\calVe}$
have been derived elsewhere \cite{FL-PLA,L-Magdy} in detail.  After assuming that the fields decay far away from their
sources sufficiently rapidly, those statements deliver
\begin{eqnarray}
\nonumber
&&
\displaystyle{
\left.\begin{array}{c}
\#E(\#r) 
\\[5pt]
\#0
\end{array}\right\}
=\#E\source(\#r)+
\int\limits_{\calS}\lec\=G^{ee}(\#r,\#r^\prime)\.
\les\un(\#r^\prime)\times\#H\plus(\#r^\prime)\ris
\right.
}
\\[10pt]
&&\qquad 
\left.-\=G^{em}(\#r,\#r^\prime)\.
\les\un(\#r^\prime)\times\#E\plus(\#r^\prime)\ris
\ric\,d^2\#r^\prime\,,\qquad
\displaystyle{
\left\{\begin{array}{l}
\#r\in{\calVe}
\\[5pt]
\#r\in{\calVi}
\end{array}\right.,
}
\l{newE64}
\end{eqnarray}
and
\begin{eqnarray}
\nonumber
&&
\displaystyle{
\left.\begin{array}{c}
\#H(\#r) 
\\[5pt]
\#0
\end{array}\right\}
=\#H\source(\#r)+
\int\limits_{\calS}\lec
\=G^{me}(\#r,\#r^\prime)\.
\les\un(\#r^\prime)\times\#H\plus(\#r^\prime)\ris 
\right.
}
\\[10pt]
&&\qquad 
\left.-\=G^{mm}(\#r,\#r^\prime)\.
\les\un(\#r^\prime)\times\#E\plus(\#r^\prime)\ris
\ric\,d^2\#r^\prime\,,\quad
\displaystyle{
\left\{\begin{array}{l}
\#r\in{\calVe}
\\[5pt]
\#r\in{\calVi}
\end{array}\right..
}
\l{newH64}
\end{eqnarray}
In these equations, $\un(\#r)$ is the unit normal to $\calS$ pointing into ${\calVe}$ (see Fig.~\ref{Fig1}),
whereas $\#E\plus(\#r)$ and $\#H\plus(\#r)$ are the electric and magnetic field
phasors on the \textit{external} side of $\calS$.

When $\#r\in{\calVi}$, Eqs.~\r{newE64} and \r{newH64} constitute the Ewald--Oseen
extinction theorem \cite{L-Magdy}. With $\#r$ chosen on the exterior side of $\calS$,  Eqs.~\r{newE64} and \r{newH64}
are {\it surface integral equations} because the unknown quantities,
 $\#E (\#r)$ and $\#H (\#r)$, appear inside as well as outside the surface integrals.

\section{Two new relations}\label{newrels}

Let us revert to Sec.~\ref{vie} and focus on $\#r\in\calVe$. Then,  the \textit{scattered} field phasors
can be identified as
\begin{equation}
\left.\begin{array}{l}
\#E\sca(\#r)=\#E(\#r)-\#E\source(\#r) 
\\[5pt]
\#H\sca(\#r)=\#H(\#r)-\#H\source(\#r) 
\end{array}
\right\}\,,\quad \#r\in\calVe\,.
\label{GeorGmor-sca}
\end{equation}
Furthermore, the actual field phasors at $\#r\in\calVi$ must be identified as
the \textit{internal} field phasors, i.e.,
\begin{equation}
\left.\begin{array}{l}
\#E\inside(\#r)=\#E(\#r) 
\\[5pt]
\#H\inside(\#r)=\#H(\#r) 
\end{array}
\right\}\,,\quad \#r\in\calVi\,.
\label{GeorGmor-int}
\end{equation}
Accordingly, 
Eqs.~\r{GeorGmor-eb}
and \r{GeorGmor-mb} can be recast as
\begin{eqnarray}
 \nonumber
&&
\displaystyle{
\#E\sca(\#r)= -i\omega\int\limits_{{{\calVi}}} 
\lec \=G^{ee}(\#r,\#r^\prime)\.\les \nueep\.\#E\inside(\#r^\prime) +\nuemp\.\#H\inside(\#r^\prime)\ris
\ric d^3\#r^\prime
}
\\[5pt]
&&
\displaystyle{
-i\omega\int\limits_{{{\calVi}}}
\lec
\=G^{em}(\#r,\#r^\prime)\. \les\numep\.\#E\inside(\#r^\prime) + \nummp\.\#H\inside(\#r^\prime)\ris
\ric d^3\#r^\prime
}\,,
\nonumber
\\[5pt]
&&\qquad\qquad \#r\in\calVe\,,
\label{GeorGmor-esca}
\end{eqnarray}
and
\begin{eqnarray}
 \nonumber
&&
\displaystyle{
\#H\sca (\#r)=  -i\omega\int\limits_{{{\calVi}}}
\lec \=G^{me}(\#r,\#r^\prime)\.\les \nueep\.\#E\inside(\#r^\prime) +\nuemp\.\#H\inside(\#r^\prime)\ris
\ric d^3\#r^\prime
}
\\[5pt]
&&
\displaystyle{
-i\omega\int\limits_{{{\calVi}}}
\lec
\=G^{mm}(\#r,\#r^\prime)\. \les\numep\.\#E\inside(\#r^\prime) + \nummp\.\#H\inside(\#r^\prime)\ris
\ric d^3\#r^\prime
}\,.
\nonumber
\\[5pt]
&&\qquad\qquad \#r\in\calVe\,,
\label{GeorGmor-msca}
\end{eqnarray}
respectively.

Next, let us revert to Sec.~\ref{sie} and also focus on $\#r\in\calVe$ but with the additional
stipulation that $\#r\notin\calS$. Equations~\r{newE64},
\r{newH64}, and \r{GeorGmor-sca}  then yield
\begin{eqnarray}
\nonumber
&&
\displaystyle{
\#E\sca(\#r)
=\int\limits_{\calS}\lec\=G^{ee}(\#r,\#r^\prime)\.
\les\un(\#r^\prime)\times\#H\plus(\#r^\prime)\ris
\right.
}
\\[10pt]
&&\qquad 
\left.-\=G^{em}(\#r,\#r^\prime)\.
\les\un(\#r^\prime)\times\#E\plus(\#r^\prime)\ris
\ric\,d^2\#r^\prime\,,\qquad
\displaystyle{
\#r\in{\calVe-\calS}
}\,,
\l{newE64a}
\end{eqnarray}
and
\begin{eqnarray}
\nonumber
&&
\displaystyle{
\#H\sca(\#r)=
\int\limits_{\calS}\lec
\=G^{me}(\#r,\#r^\prime)\.
\les\un(\#r^\prime)\times\#H\plus(\#r^\prime)\ris 
\right.
}
\\[10pt]
&&\qquad 
\left.-\=G^{mm}(\#r,\#r^\prime)\.
\les\un(\#r^\prime)\times\#E\plus(\#r^\prime)\ris
\ric\,d^2\#r^\prime\,,\quad
\displaystyle{
\#r\in{\calVe-\calS}
}\,.
\l{newH64a}
\end{eqnarray} 
With $\#E\minus(\#r)$ and $\#H\minus(\#r)$ denoting the electric and magnetic field
phasors on the \textit{internal} side of $\calS$, the standard boundary conditions
\begin{equation}
\left.\begin{array}{l}
\un(\#r)\times\#E\plus(\#r)= \un(\#r)\times\#E\minus(\#r)
\\[5pt]
 \un(\#r)\times\#H\plus(\#r)= \un(\#r)\times\#H\minus(\#r)
\end{array}
\right\}\,,\quad \#r\in\calS\,,
\end{equation}
can be restated as
\begin{equation}
\left.\begin{array}{l}
\un(\#r)\times\#E\plus(\#r)= \un(\#r)\times\#E\inside(\#r)
\\[5pt]
 \un(\#r)\times\#H\plus(\#r)= \un(\#r)\times\#H\inside(\#r)
\end{array}
\right\}\,,\quad \#r\in\calS\,.
\end{equation}
Accordingly, Eqs.~\r{newE64a} and \r{newH64a} can be rewritten as
\begin{eqnarray}
\nonumber
&&
\displaystyle{
\#E\sca(\#r)
=\int\limits_{\calS}\lec\=G^{ee}(\#r,\#r^\prime)\.
\les\un(\#r^\prime)\times\#H\inside(\#r^\prime)\ris
\right.
}
\\[10pt]
&&\qquad 
\left.-\=G^{em}(\#r,\#r^\prime)\.
\les\un(\#r^\prime)\times\#E\inside(\#r^\prime)\ris
\ric\,d^2\#r^\prime\,,\qquad
\displaystyle{
\#r\in{\calVe-\calS}
}\,,
\l{newE64b}
\end{eqnarray}
and
\begin{eqnarray}
\nonumber
&&
\displaystyle{
\#H\sca(\#r)=
\int\limits_{\calS}\lec
\=G^{me}(\#r,\#r^\prime)\.
\les\un(\#r^\prime)\times\#H\inside(\#r^\prime)\ris 
\right.
}
\\[10pt]
&&\qquad 
\left.-\=G^{mm}(\#r,\#r^\prime)\.
\les\un(\#r^\prime)\times\#E\inside(\#r^\prime)\ris
\ric\,d^2\#r^\prime\,,\quad
\displaystyle{
\#r\in{\calVe-\calS}
}\,,
\l{newH64b}
\end{eqnarray} 
respectively. According to these equations, the tangential components of the internal electric and magnetic field
phasors on the object's surface are \textit{equivalent} to surface current densities on $\calS$.

On comparing Eqs.~\r{GeorGmor-esca} and \r{newE64b}, we get
\begin{eqnarray}
 \nonumber
&&
\displaystyle{
\int\limits_{{{\calVi}}} 
\lec \=G^{ee}(\#r,\#r^\prime)\.\les \nueep\.\#E\inside(\#r^\prime) +\nuemp\.\#H\inside(\#r^\prime)\ris
\ric d^3\#r^\prime
}
\\[5pt]
&&\qquad+
\displaystyle{
\int\limits_{{{\calVi}}}
\lec
\=G^{em}(\#r,\#r^\prime)\. \les\numep\.\#E\inside(\#r^\prime) + \nummp\.\#H\inside(\#r^\prime)\ris
\ric d^3\#r^\prime
}
\nonumber
\\[5pt]
&&\qquad\qquad
\nonumber
\displaystyle{
=\frac{i}{\omega}\int\limits_{\calS}\lec\=G^{ee}(\#r,\#r^\prime)\.
\les\un(\#r^\prime)\times\#H\inside(\#r^\prime)\ris
\right.
}
\\[10pt]
&&\qquad \qquad\qquad
\left.-\=G^{em}(\#r,\#r^\prime)\.
\les\un(\#r^\prime)\times\#E\inside(\#r^\prime)\ris
\ric\,d^2\#r^\prime\,,\qquad
\displaystyle{
\#r\in{\calVe-\calS}
}\,.
\label{newrel-e}
\end{eqnarray}
Likewise, Eqs.~\r{GeorGmor-msca} and \r{newH64b} deliver the relation
\begin{eqnarray}
 \nonumber
&&
\displaystyle{
\int\limits_{{{\calVi}}} 
\lec \=G^{me}(\#r,\#r^\prime)\.\les \nueep\.\#E\inside(\#r^\prime) +\nuemp\.\#H\inside(\#r^\prime)\ris
\ric d^3\#r^\prime
}
\\[5pt]
&&\qquad+
\displaystyle{
\int\limits_{{{\calVi}}}
\lec
\=G^{mm}(\#r,\#r^\prime)\. \les\numep\.\#E\inside(\#r^\prime) + \nummp\.\#H\inside(\#r^\prime)\ris
\ric d^3\#r^\prime
}
\nonumber
\\[5pt]
&&\qquad\qquad
\nonumber
\displaystyle{
=\frac{i}{\omega}\int\limits_{\calS}\lec\=G^{me}(\#r,\#r^\prime)\.
\les\un(\#r^\prime)\times\#H\inside(\#r^\prime)\ris
\right.
}
\\[10pt]
&&\qquad \qquad\qquad
\left.-\=G^{mm}(\#r,\#r^\prime)\.
\les\un(\#r^\prime)\times\#E\inside(\#r^\prime)\ris
\ric\,d^2\#r^\prime\,,\qquad
\displaystyle{
\#r\in{\calVe-\calS}
}\,.
\label{newrel-m}
\end{eqnarray}

Equations~\r{newrel-e} and \r{newrel-m} underwrite the novelty of this paper. Since both equations can be derived from each other
because
\begin{equation}
\left.\begin{array}{l}
\lplus\.\#H\sca(\#r)+i\omega
\=\eps \.\#E\sca(\#r)=\#0 
\\[5pt]
\lminus\.\#E\sca(\#r)-i\omega
\=\mu \.\#H\sca(\#r)
=\#0  
\end{array}\right\}\,,\qquad\#r\in \calVe\,,
\label{M1worM2wor-sca}
\end{equation}
they are not independent. However, one may be easier than the other to use in some situations.
Also, note that Eq.~\r{newrel-e} does not change if both sides of it are operated on from the left 
 by $\left(\lplus\.\=\mu^{-1}\.\lminus\right)\.$, as shown in the Appendix. Likewise,
Eq.~\r{newrel-m} does not change if both sides of it are operated on from the left by $\left(\lminus\.\=\eps^{-1}\.\lplus\right)\.$.

{Parenthetically, Eqs.~\r{newrel-e} and \r{newrel-m} arise because the scattered field phasors must be
the same whether derived from surface integral equations or from volume integral equations, both types
of integral equations being exact and derivable from the frequency-domain Maxwell curl postulates.}

\subsection{Alternative forms}
Equations~\r{newrel-e} and \r{newrel-m}  can be put in  alternative forms
using Eqs.~\r{DGF-symm}. For example,
Eq.~\r{newrel-e} may be rewritten as either 
\begin{eqnarray}
 \nonumber
&&
\displaystyle{
\int\limits_{{{\calVi}}} 
 \exp\les-2i\omega\#\Gamma\. \#r^\prime \ris
\lec \les \nueep\.\#E\inside(\#r^\prime) +\nuemp\.\#H\inside(\#r^\prime)\ris
\.\=G^{ee}(\#r^\prime,\#r)
\ric d^3\#r^\prime
}
\\[5pt]
&&\qquad-
\displaystyle{
\int\limits_{{{\calVi}}}
\exp\les-2i\omega\#\Gamma\. \#r^\prime \ris
\lec
 \les\numep\.\#E\inside(\#r^\prime) + \nummp\.\#H\inside(\#r^\prime)\ris
 \.\=G^{me}(\#r^\prime,\#r) 
\ric d^3\#r^\prime
}
\nonumber
\\[5pt]
&&\qquad\qquad
\nonumber
\displaystyle{
=\frac{i}{\omega} \exp\les-2i\omega\#\Gamma\. \#r \ris
\int\limits_{\calS}\lec\=G^{ee}(\#r,\#r^\prime)\.
\les\un(\#r^\prime)\times\#H\inside(\#r^\prime)\ris
\right.
}
\\[10pt]
&&\qquad \qquad\qquad
\left.-\=G^{em}(\#r,\#r^\prime)\.
\les\un(\#r^\prime)\times\#E\inside(\#r^\prime)\ris
\ric\,d^2\#r^\prime\,,\qquad
\displaystyle{
\#r\in{\calVe}
}\,,
\label{newrel-e-1}
\end{eqnarray}
or
\begin{eqnarray}
 \nonumber
&&
\displaystyle{
\int\limits_{{{\calVi}}} 
\exp\les-2i\omega\#\Gamma\. \#r^\prime \ris
\lec \les \nueep\.\#E\inside(\#r^\prime) +\nuemp\.\#H\inside(\#r^\prime)\ris
\.\=G^{ee}(\#r^\prime,\#r) 
\ric d^3\#r^\prime
}
\\[5pt]
&&\qquad-
\displaystyle{
\int\limits_{{{\calVi}}}
\exp\les-2i\omega\#\Gamma\. \#r^\prime \ris
\lec
 \les\numep\.\#E\inside(\#r^\prime) + \nummp\.\#H\inside(\#r^\prime)\ris
 \.\=G^{me}(\#r^\prime,\#r) 
\ric d^3\#r^\prime
}
\nonumber
\\[5pt]
&&\qquad\qquad
\nonumber
\displaystyle{
=\frac{i}{\omega}\int\limits_{\calS}
\exp\les-2i\omega\#\Gamma\. \#r^\prime \ris\lec 
\les\un(\#r^\prime)\times\#H\inside(\#r^\prime)\ris
\.\=G^{ee}(\#r^\prime,\#r)  
\right.
}
\\[10pt]
&&\qquad \qquad\qquad
\left.+ 
\les\un(\#r^\prime)\times\#E\inside(\#r^\prime)\ris
 \.\=G^{me}(\#r^\prime,\#r)  
\ric\,d^2\#r^\prime\,,\qquad
\displaystyle{
\#r\in{\calVe}
}\,.
\label{newrel-e-2}
\end{eqnarray}

\subsection{Scattering in free space}
When $\calVe$ is vacuous, $\=\eps=\epso\,\=I$, $\=\mu=\muo\,\=I$, $\=\xi=\=0$, and
$\#K=\#\Gamma=\#0$. As a result,

\begin{equation}
\left.\begin{array}{l}
\=G^{ee} (\#r,\#r^\prime)= i\omega\muo\, \=G_{0} (\#r,\#r^\prime)
\\[5pt]
\=G^{mm} (\#r,\#r^\prime)= i\omega\epso\, \=G_{0} (\#r,\#r^\prime)
\\[8pt]
\=G^{em}(\#r,\#r^\prime)=-\=G^{me}(\#r,\#r^\prime)=- \nabla \times \=G_{0} (\#r,\#r^\prime)
\end{array}
\right\}\,,
\end{equation}
where
\begin{equation}
\=G_{0} (\#r,\#r^\prime)= \left(\=I + \ko^{-2}\nabla\nabla\right)
\frac{\exp\left(i\ko\vert\#r-\#r^\prime\vert\right)} {4\pi\vert\#r-\#r^\prime\vert}
\end{equation}
is the usual dyadic Green function for free space and $\ko=\omega\sqrt{\epso\muo}$.
Equation~\r{newrel-e} then simplifies to
\begin{eqnarray}
 \nonumber
&&
\displaystyle{i\omega\muo
\int\limits_{{{\calVi}}} 
\lec \=G_{0}(\#r,\#r^\prime)\.\les \nueep\.\#E\inside(\#r^\prime) +\nuemp\.\#H\inside(\#r^\prime)\ris
\ric d^3\#r^\prime
}
\\[5pt]
&&\qquad-
\displaystyle{\nabla\times
\int\limits_{{{\calVi}}}
\lec
\=G_{0}(\#r,\#r^\prime)\. \les\numep\.\#E\inside(\#r^\prime) + \nummp\.\#H\inside(\#r^\prime)\ris
\ric d^3\#r^\prime
}
\nonumber
\\[5pt]
&&\qquad\qquad
\nonumber
\displaystyle{
=-\muo\int\limits_{\calS}\lec  \=G_{0}(\#r,\#r^\prime)\.
\les\un(\#r^\prime)\times\#H\inside(\#r^\prime)\ris
\ric\,d^2\#r^\prime
}
\\[10pt]
&&\qquad \qquad 
+\frac{i}{\omega}\nabla\times
\int\limits_{\calS}\lec
\=G_{0}(\#r,\#r^\prime)\.
\les\un(\#r^\prime)\times\#E\inside(\#r^\prime)\ris
\ric\,d^2\#r^\prime\,,\quad
\displaystyle{
\#r\in{\calVe-\calS}
}\,,
\label{newrel-e-fs}
\end{eqnarray}
and Eq.~\r{newrel-m} to
\begin{eqnarray}
 \nonumber
&&
\displaystyle{\nabla\times
\int\limits_{{{\calVi}}} 
\lec \=G_{0}(\#r,\#r^\prime)\.\les \nueep\.\#E\inside(\#r^\prime) +\nuemp\.\#H\inside(\#r^\prime)\ris
\ric d^3\#r^\prime
}
\\[5pt]
&&\qquad+
\displaystyle{i\omega\epso
\int\limits_{{{\calVi}}}
\lec
\=G_{0}(\#r,\#r^\prime)\. \les\numep\.\#E\inside(\#r^\prime) + \nummp\.\#H\inside(\#r^\prime)\ris
\ric d^3\#r^\prime
}
\nonumber
\\[5pt]
&&\qquad\qquad
\nonumber
\displaystyle{
=\frac{i}{\omega}\nabla\times\int\limits_{\calS}\lec \=G_{0}(\#r,\#r^\prime)\.
\les\un(\#r^\prime)\times\#H\inside(\#r^\prime)\ris
\ric\,d^2\#r^\prime
}
\\[10pt]
&&\qquad \qquad
+\epso\int\limits_{\calS}\lec\=G_{0}(\#r,\#r^\prime)\.
\les\un(\#r^\prime)\times\#E\inside(\#r^\prime)\ris
\ric\,d^2\#r^\prime\,,\quad
\displaystyle{
\#r\in{\calVe-\calS}
}\,.
\label{newrel-m-fs}
\end{eqnarray}

If we additionally suppose that $\calVi$ is occupied by a homogeneous, isotropic dielectric medium,
then $\nueep=\nu_{ee}\=I$, $\nuemp= \=0$, $\numep= \=0$, and $\nummp= \=0$ $\forall\,\#r^\prime\in\calVi$.
Accordingly, Eq.~\r{newrel-e-fs} simplifies to
\begin{eqnarray}
 \nonumber
&&
\displaystyle{\omega^2\muo\nu_{ee}
\int\limits_{{{\calVi}}} 
\les \=G_{0}(\#r,\#r^\prime)\.\#E\inside(\#r^\prime)  
\ris d^3\#r^\prime
}
\\[5pt]
&&\qquad\qquad
\nonumber
\displaystyle{
=i\omega\muo\int\limits_{\calS}\lec \=G_{0}(\#r,\#r^\prime)\.
\les\un(\#r^\prime)\times\#H\inside(\#r^\prime)\ris
\ric\,d^2\#r^\prime
}
\\[10pt]
&&\qquad \qquad 
+\nabla\times\int\limits_{\calS}\lec\=G_{0}(\#r,\#r^\prime)\.
\les\un(\#r^\prime)\times\#E\inside(\#r^\prime)\ris
\ric\,d^2\#r^\prime\,,\quad
\displaystyle{
\#r\in{\calVe-\calS}
}\,,
\label{newrel-e-fs-diel}
\end{eqnarray}
and Eq.~\r{newrel-m-fs} to
\begin{eqnarray}
 \nonumber
&&
\displaystyle{\nu_{ee}\nabla\times
\int\limits_{{{\calVi}}} 
\les \=G_{0}(\#r,\#r^\prime)\.\#E\inside(\#r^\prime)  
\ris d^3\#r^\prime
}
\\[5pt]
&&\qquad\qquad
\nonumber
\displaystyle{
=\frac{i}{\omega}\int\limits_{\calS}\lec \nabla\times\=G_{0}(\#r,\#r^\prime)\.
\les\un(\#r^\prime)\times\#H\inside(\#r^\prime)\ris
\ric\,d^2\#r^\prime
}
\\[10pt]
&&\qquad \qquad
+\epso\int\limits_{\calS}\lec \=G_{0}(\#r,\#r^\prime)\.
\les\un(\#r^\prime)\times\#E\inside(\#r^\prime)\ris
\ric\,d^2\#r^\prime\,,\quad
\displaystyle{
\#r\in{\calVe-\calS}
}\,.
\label{newrel-m-fs-diel}
\end{eqnarray}
Equation~\r{newrel-e-fs} can also be derived  \cite{Berg} using the second vector-dyadic Green theorem \cite[p.~300]{Tai}.
{It is a simple matter to apply Eqs.~\r{newrel-e-fs-diel} and \r{newrel-m-fs-diel} to multipolar scattering by an isotropic dielectric sphere and thus validate the analytical Lorenz--Mie theory \cite{Lorenz,Mie,BH83}.}

\section{New principle}\label{newprin}
Equations~\r{newrel-e} and \r{newrel-m} allow us to enunciate a new principle for scattering inside
a Huygens bianisotropic medium. 
The left sides of both equations contain the excess field phasors ($\#D\xs$ and $\#B\xs$) inside
the scattering object,  {as defined in Eqs.~(\ref{excess}).} The right sides of both equations contain the tangential components of
the internal field phasors ($\#E\inside$ and $\#H\inside$) on the surface of the same object. The new principle
may be enunciated as follows:

\textit{The excess field phasors inside a bounded 3D scattering object 
act as volume current densities, and the tangential components of the internal field phasors on the
surface of the same object act as surface current densities, to radiate identical field phasors in the external region, provided the external medium is a Huygens bianisotropic medium.}

The commonest Huygens bianisotropic medium is   free space. Thus, this new principle certainly applies
to scattering in free space, and it can be  useful in checking the results of semi-analytical and numerical 
methods for solving  scattering problems in free space \cite{Kahnert},  {so long as  the scattering object is composed of a linear medium.} 

 {In closing, the new principle is exact, just like the celebrated optical theorem  \cite{deHoop,Newton} is for the scattering of
a plane wave by an object composed of a linear medium and surrounded by free space. Unlike the optical theorem however, the new principle applies not only to incident plane waves but to any incident time-harmonic field. 
Next, the optical theorem applies to power density, but the new principle to the field phasors.
Also, whereas the optical theorem has only been extended to the external medium being an isotropic chiral medium 
\cite[Sec.~5-2.7]{Beltrami}, Sec.~\ref{newrels} shows that the new principle applies when the external medium is far more general than an isotropic chiral medium.  Thus, the new principle has a huge scope in computational electromagnetics for validation of scattering results provided by diverse semi-analytical and numerical methods.}

\section*{Appendix}\label{App}
\setcounter{equation}{0}
\makeatletter 
\renewcommand{\theequation}{A\@arabic\c@equation}
\makeatother

The integrands on both sides of Eq.~\r{newrel-e} are of the form
\begin{equation}
\=G^{ee}(\#r,\#r^\prime)\.\#X(\#r^\prime)+\=G^{em}(\#r,\#r^\prime)\.\#Y(\#r^\prime)\,,
\end{equation}
with $\#r$ not lying in the integration domain. 
Therefore,
\begin{eqnarray}
\nonumber
&&\left(\lplus\.\=\mu^{-1}\.\lminus\right)\.\=G^{ee}(\#r,\#r^\prime)\.\#X(\#r^\prime)
\\[5pt]
\nonumber
&&\qquad=\les\omega^2\=\eps\.\=G^{ee}(\#r,\#r^\prime)+i\omega\=I\delta(\#r-\#r^\prime)\ris\.\#X(\#r^\prime)
\\[5pt]
&&\qquad= \omega^2\=\eps\.\=G^{ee}(\#r,\#r^\prime)\.\#X(\#r^\prime)
\end{eqnarray}
follows after using Eq.~\r{six},
and
\begin{eqnarray}
\nonumber
&&\left(\lplus\.\=\mu^{-1}\.\lminus\right)\.\=G^{em}(\#r,\#r^\prime)\.\#Y(\#r^\prime)
\\[5pt]
\nonumber
&&\qquad=-\frac{1}{i\omega}
\left(\lplus\.\=\mu^{-1}\.\lminus\right)\.\=\eps^{-1}\.\lplus\.
\=G^{mm}(\#r,\#r^\prime)\.\#Y(\#r^\prime)
\\[5pt]
\nonumber
&&\qquad=-\frac{1}{i\omega}
\left(\lplus\.\=\mu^{-1}\right)\.
\les\omega^2\=\mu\.\=G^{mm}(\#r,\#r^\prime)+i\omega\=I\delta(\#r-\#r^\prime)\ris\.\#Y(\#r^\prime)
\\[5pt]
\nonumber
&&\qquad=i\omega\lplus\.\=G^{mm}(\#r,\#r^\prime)\.\#Y(\#r^\prime)
\\[5pt]
&&\qquad=\omega^2\=\eps\.\=G^{em}(\#r,\#r^\prime)\.\#Y(\#r^\prime)
 \end{eqnarray}
follows using Eqs.~\r{seven} and \r{five-em}. Accordingly,
\begin{eqnarray}
&&
\nonumber
\left(\lplus\.\=\mu^{-1}\.\lminus\right)\.
\les\=G^{ee}(\#r,\#r^\prime)\.\#X(\#r^\prime)+\=G^{em}(\#r,\#r^\prime)\.\#Y(\#r^\prime)\ris
\\[5pt]
&&\qquad=\omega^2\=\eps\.
\les\=G^{ee}(\#r,\#r^\prime)\.\#X(\#r^\prime)+\=G^{em}(\#r,\#r^\prime)\.\#Y(\#r^\prime)\ris
\end{eqnarray}
so that Eq.~\r{newrel-e} does not change if both sides of it are operated on from the left 
 by $\left(\lplus\.\=\mu^{-1}\.\lminus\right)\.$. In the same way, Eq.~\r{newrel-m} does not change if both sides of it are operated on from the left by $\left(\lminus\.\=\eps^{-1}\.\lplus\right)\.$.
 
\vspace{1cm}
\noindent \textbf{Acknowledgements.} The author thanks the Charles Godfrey Binder Endowment at Penn State
for ongoing support of his research activities.
 
\vspace{1cm}
\noindent \textbf{Competing interests.} The author has  no competing interests to declare that are relevant to the content of this paper.

 \end{document}